\documentclass[12pt]{article}
\usepackage{epsfig,psfrag,amsmath,amssymb,latexsym}
\usepackage{amscd }
\usepackage{amsfonts}
\usepackage{graphicx}
\usepackage{float}
\usepackage{epic}
\usepackage{curves}
\usepackage{bezier}
\usepackage{wrapfig}
\usepackage[T1]{fontenc}
\usepackage[dvips]{color}
\newtheorem{theorem}{Theorem}[section]

\newcommand{\A}{{\cal{A}}}

\newcommand{\p}{\bar{p}}
\newcommand{\q}{\bar{q}}
\numberwithin{equation}{section}
\begin{document}
\title{Detecting the homology of DNA-sequences based on the variety
of optimal alignments: a case study}
\author{Erik Hirmo, Jüri Lember \thanks{{Supported by the Estonian Science Foundation
Grant nr. 9288 and  by targeted financing project SF0180015s12}},
Heinrich Matzinger}

\maketitle

{\small \vspace{0cm} \hbox{\hspace{0 cm}\vbox{\noindent \hsize=5cm  University of Tartu\\
Liivi 2-513 50409, Tartu\\
Estonia\\} \vbox{\noindent \hsize=5cm  University of Tartu\\
Liivi 2-513 50409, Tartu\\
Estonia\\
E-mail: jyril@ut.ee} \vbox{\noindent \hsize=7cm
 Georgia Tech\\
School of Mathematics\\
Atlanta, Georgia 30332-0160, U.S.A.\\
E-mail: matzing@math.gatech.edu}} \vskip 1cm \hfill}
\vspace{0.5cm}\noindent
%------------
\section{Introduction}
%------------
We consider the problem of detecting the dependence/homology of two
sequences of finite alphabet with comparable length. The classical
measures of similarity in the sequence comparison  are based on the
score of  optimal alignment of the sequences of interest (see, e.g.
\cite{Chao, Durbin,Vingron, Watermanphase, watermanintrocompbio}).
The optimal alignment is in general not unique, but all optimal
alignments provide the same score. Hence, the difference between
various optimal alignments is not taken into account. Our method is
based on the observation that for many scoring schemes, especially
for the longest common subsequence (LCS) scoring, the optimal
alignments are the more different the unrelated are the sequences.
This gives the idea to use the variety of the optimal alignments as
an additional measure of the homology. A (partial) theoretical
justification of the idea is given in \cite{vollmer}, where the
differences between LCS-optimal alignments were measured in terms of
distances between so-called extremal alignments. The main result of
\cite{vollmer} states that for related sequences (in certain sense),
the distance between extremal is of order $\ln n$, where $n$ is the
length of the both sequences. It has not proven that for independent
sequences the distance between extremal alignments increases faster
than $\ln n$, but the simulations in \cite{vollmer} show that this
is indeed the case. Hence, the  distance between extremal alignments
could provide important information about the similarity of the
sequences. To test that idea for actual DNA-sequences, in this note
we apply the above mentioned ideas to four biologically similar
genes, and we compare our test results with the outcomes of commonly
used BLAST program (see \cite{BLAST, Chao}).\\\\
The paper is organized as follows. In the next section, we briefly
explain the setup and main results of \cite{vollmer}. In the last
section, the results of the simulation study are presented.
%-------------
\section{Theoretical background}
\subsection{Extremal alignments}
Let ${\cal A}$ be a finite alphabet. In the context of
DNA-sequences, ${\cal A}$ obviously  consists of four letters. In
everything that follows, $X=X_1\ldots X_n\in {\cal A}^n$ and
$Y=Y_1\ldots Y_n\in {\cal A}^n$ are two strings of length $n$. A
common subsequence of $X$ and $Y$ is a sequence that is a
subsequence of $X$ and at the same time of $Y$. We denote by $L_n$
the length of the longest common subsequence (LCS) of $X$ and $Y$.
The length of LCS is clearly an important tool in the sequence
comparison, the bigger is $L_n$ the more the sequences are
presupposed to be related.  It is well known that for ergodic
sequences (in particular, for independent $iid$ sequences), the
relative length of the LCS converges to a constant, i.e.
\begin{equation}\label{gamma}
{L_n\over n}\to \gamma,\quad\rm{ a.s.},
\end{equation}
where $\gamma$ is so-called Chvatal-Sankov constant. Unfortunately,
the constant $\gamma$ is not exactly known for as simple cases as
i.i.d. Bernoulli sequences. Moreover, the speed of convergence or
the corresponding central limit law is not known. This all makes it
very difficult to use $L_n$ for testing the independence and
motivates us to look for some alternative LCS-related criterions.
\\\\
%------
We now explain the idea of extremal alignment. Recall that $X$ and
$Y$ are both sequences of length $n$. Let there exist two subsets of
indices $\{i_1,\ldots,i_k\}\subset \{1,\ldots,n\}$ and
$\{j_1,\ldots,j_k\}\subset \{1,\ldots,n_\}$ satisfying
$i_1<i_2<\ldots <i_k$, $j_1<j_2<\ldots <j_k$ and $X_{i_1}=Y_{j_1},
X_{i_2}=Y_{j_2},\ldots, X_{i_k}=Y_{j_k}.$ Then $X_{i_1}\cdots
X_{i_k}$ is a common subsequence of $X$ and $Y$ and the pairs
\begin{equation}\label{alignment}
\{(i_1,j_1),\ldots, (i_k,j_k)\}
\end{equation}
 are  the corresponding alignment.
Then $L_n$ is  the biggest $k$ such that there exist such subsets of
indices and any alignment corresponding to a longest common
subsequence is called optimal. We consider every (optimal) alignment
(\ref{alignment}) as  a set of points in $\{1,\ldots,n\}\times
\{1,\ldots,n\}$ and we shall call it as the two-dimensional
representation of (optimal) alignment. Note that different optimal
alignments can result the same common subsequence, and usually there
are more then one common subsequence. Hence, typically,  there are
many optimal alignments represented in  $\{1,\ldots,n\}\times
\{1,\ldots,n\}$. We believe that the notions of extremal alignments
will be clear through the following example.\\\\
{\bf Example:} Let $X=$ATAGCGT, $Y=$CAACATG. There are two longest
common subsequences: {AACG} and {AACT}. Thus $L_7=4$. To every
longest common subsequence corresponds two alignment in form
(\ref{alignment}): the alignments $(1,2), (2,3), (5,5), (6,7)$ and
$(1,2), (2,3), (5,4), (6,7)$ correspond to AACG; the alignments
$(1,2),(3,3),(5,4),(7,6)$ and $(1,2),(3,3),(4,4),(7,6)$ correspond
to AACT. The corresponding two dimensional graphs are\\ \centerline{
{\footnotesize \hbox{{$\begin{array}{c|c|c|c|c|c|c|c} G& & & & & &*&
\\\hline T& & & & & & &
\\\hline A& & & & & & &\\\hline C& & & & &* & &\\\hline A& & &* & & &
&
\\\hline A&*& & & & & & \\\hline
C& & & & & &  & \\\hline
  &A&T&A&C&C&G&T
\end{array}$} {$\begin{array}{c|c|c|c|c|c|c|c} G& & & & & &*&
\\\hline T& & & & & & &
\\\hline A& & & & & & &\\\hline C& & & & *& & &\\\hline A& & &* & & &
&
\\\hline A&*& & & & & & \\\hline
C& & & & & &  & \\\hline
  &A&T&A&C&C&G&T
\end{array}$}}
\hbox{{$\begin{array}{c|c|c|c|c|c|c|c} G& & & & & & &
\\\hline T& & & & & & & *
\\\hline A& & & & & & &\\\hline C& & & & &* & &\\\hline A& & &* & & &
&
\\\hline A&*& & & & & & \\\hline
C& & & & & &  & \\\hline
  &A&T&A&C&C&G&T
\end{array}$}{$\begin{array}{c|c|c|c|c|c|c|c} G& & & & & & &
\\\hline T& & & & & & &*
\\\hline A& & & & & & &\\\hline C& & & & *& & &\\\hline A& & &* & & &
&
\\\hline A&*& & & & & & \\\hline
C& & & & & &  & \\\hline
  &A&T&A&C&C&G&T
\end{array}$}}}}
\vskip 1\baselineskip\noindent
 Putting all four alignment into one graph, we see that on some
regions all alignments are unique, but on some region, they vary:
\\\\{\small
$$\begin{array}{c|c|c|c|c|c|c|c} G& & & & & &\textcolor{red}{*}&
\\\hline T& & & & & & &\textcolor{blue}{*}
\\\hline A& & & & & & &\\\hline C& & & & \textcolor{red}{*}& \textcolor{blue}{*}& &\\
\hline A& & &* & & & &
\\\hline A&*& & & & & & \\\hline
C& & & & & &  & \\\hline
  &A&T&A&C&C&G&T
\end{array}$$}
In the picture above, the two black dots and the red dots correspond
to the alignment that lies above all others. This alignment will be
called {\it highest alignment}. Similarly the two black dots and the
blue dots correspond to the alignment that lies below all others.
This alignment will be called {\it lowest alignment}. In our
example, thus the alignment $(1,2), (2,3), (5,4), (6,7)$
(corresponding to AACG) is the highest and the alignment
$(1,2),(3,3),(5,4),(7,6)$ (corresponding to AACT) is the lowest. The
highest and lowest alignment will be called {\it extremal
alignments}.\\\\
%----------
Thus, the highest (lowest) alignment is the one that lies above
(below) all other alignments in two-dimensional representation. The
formal definition of the extremal alignments as well as the proof
the definition is correct can be found in \cite{vollmer}. For big
$n$, we usually align the dots in the two dimensional representation
by lines. In Figure 1, taken from \cite{vollmer}, there are extremal
alignments (red) of two independent {\it iid} sequences of length
$n=1000$. It is visible that the extremal alignments are rather far
from each other, in particular, the maximum vertical and horizontal
distances are relatively big.
%-------
\begin{figure}[H]
\includegraphics[height=10cm]{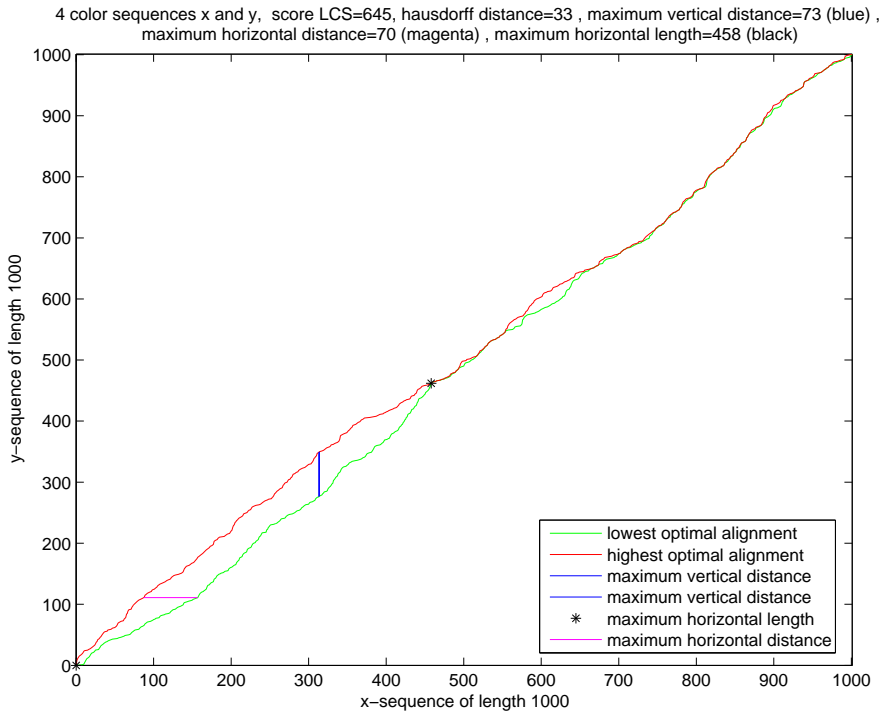}
\caption{The extremal alignments of two independent iid four letter
sequences}
\end{figure}
%-----------
\subsection{Related sequences}
%---------
Unrelated sequences $X$ and $Y$ are independent. In our setup, the
relatedness is based on the assumption that there exists a common
ancestor, from which both sequences $X$ and $Y$ are obtained by
independent random mutations and deletions. In the following, the
common ancestor is an ${\cal A}$-valued {\it iid} process
$Z_1,Z_2,\ldots$. A letter $Z_i$ has a probability to mutate
according to a transition matrix that does not depend on $i$. Hence,
a mutation of the letter $Z_i$ can be formalized as $f(Z_i, \xi_i)$,
where $f: \A \times \mathbb{R}\to \A$ is a mapping and $\xi_i$ is a
 uniformly distributed random variable. The mapping $f_i(\cdot):=f(\cdot,
\xi_i)$ from $\A$ to $\A$ will be referred as the random mapping.
The mutations of the letters are assumed to be independent. This
means that the random variables $\xi_1,\xi_2,\ldots$ or the random
mappings $f_1,f_2,\ldots$ are independent (and identically
distributed). After mutations, the sequence is
$f_1(Z_1),f_2(Z_2),\ldots.$ Some of its elements disappear. This is
modeled via a deletion process $D^x_1,D^x_2,\ldots$ that is assumed
to be an {\it iid} Bernoulli sequence with parameter $p$ i.e.
$P(D^x_i=1)=p$. If $D^x_i=0$, then $f_i(Z_i)$ is deleted. The
resulting sequence, let it be $X$, is, therefore, the following:
$X_i=f_j(Z_j)$ if and only if $D^x_j=1$ and $\sum_{k=1}^{j}D^x_k=i$.
Similarly, the sequence $Y$ is obtained from $Z$. For mutations, fix
an  {\it iid} uniformly distributed sequence $\eta_1,\eta_2,\ldots$
so that the mutated sequence is $ h_1(Z_1), h_2(Z_2),\ldots$ with $
h_i(\cdot):=f(\cdot, \eta_i).$ Note that the transition matrix
corresponding to $Y$-mutations equals the one corresponding to
$X$-mutations implying that the random mappings $h_i$ and $f_i$ have
the same distribution. Since the mutations of $X$ and $Y$ are
supposed to be independent, we assume the sequences $\xi$ and $\eta$
or the random mappings sequences $f_1, f_2,\ldots $ and
$h_1,h_2,\ldots$ are independent. Note that then the pairs
$(f_1(Z_1),h_1(Z_1)),(f_2(Z_2),h_2(Z_2)),\ldots$ are independent,
but $f_i(Z_i)$ and $h_i(Z_i)$, in general, are not. Finally, some of
the elements of $h_1(Z_1),h_2(Z_2),\ldots$ are deleted according to
a deletion process  $D^y_1,D^y_2,\ldots$ consisting of ${\it iid}$
Bernoulli random variables with the same parameter as $D^x$ but
independent of $D^x.$   The remaining elements define $Y$-sequence.
Note that our definition of relatedness
involves the independent sequences as a special case, when the functions $f$ does not depend on $Z$. \\\\
{\bf Example:} The following table illustrates the generic process
of obtaining $X$ and $Y$. {\small
$$\begin{array}{c|c|c|c|c|c|c|c} Z:&Z_1&Z_2&Z_3&Z_4&Z_5&Z_6&\text{common
ancestor}\\\hline\\\hline
f(Z):&f_1(Z_1)&f_2(Z_2)&f_3(Z_3)&f_4(Z_4)&f_5(Z_5)&f_6(Z_6)&\text{X
mutations}\\\hline D^x:& 0&1&1&0&0&1&\text{X deletions}\\\hline X:&
&X_1&X_2& & &X_3& \\\hline\\\hline
h(Z):&h_1(Z_1)&h_2(Z_2)&h_3(Z_3)&h_4(Z_4)&h_5(Z_5)&h_6(Z_6)&\text{Y
mutations}\\\hline D^y:& 1&1&1&0&1&0&\text{Y deletions}\\\hline Y:&
Y_1&Y_2 &Y_3& &Y_4& &
\end{array}$$}
%---------
\begin{figure}[H]
\includegraphics[height=10cm]{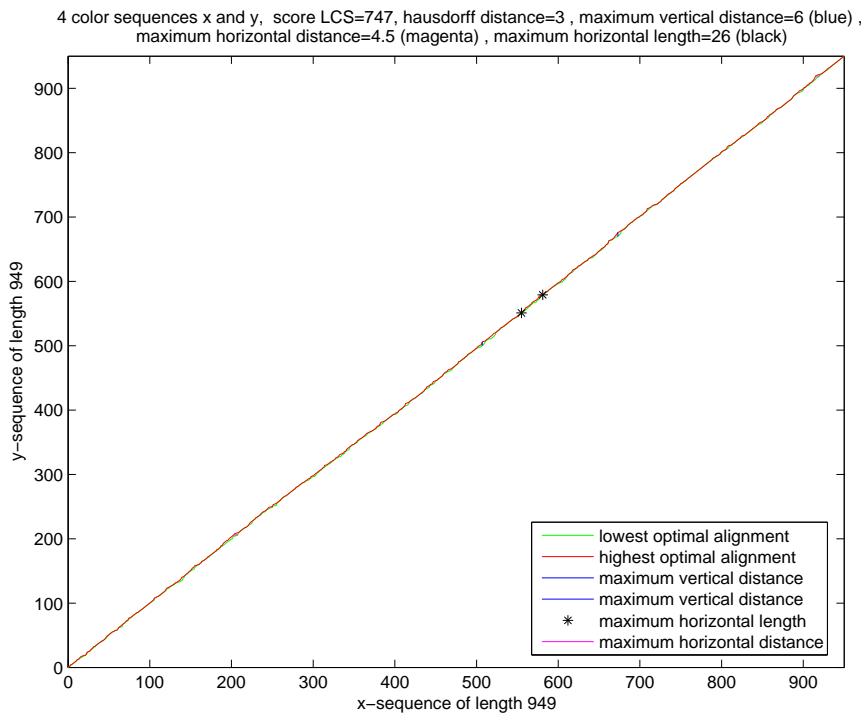}
\caption{The extremal alignments of two related four letter
sequences}
\end{figure}
%-----------
In \cite{vollmer}, the related sequences were simulated and the
corresponding extremal alignments  were found. Figure 2 presents a
typical picture or extremal alignments of two related sequences of
length 1000. Clearly the extremal alignments are close to each
other; in particular the maximal vertical and horizontal distance is
much smaller than these ones in Figure 1. The closeness of the
extremal alignments of the related sequences follows from the main
result of \cite{vollmer}. Before we state the result formally, some
notations need to be introduced. Let $\gamma_R$ be the limit of
(\ref{gamma}), where $X$ and $Y$ are related. Typically
$\gamma_R>\gamma$, where $\gamma$ is the limit of independent
sequences with the same laws. The existence of $\gamma_R$ is proven
in \cite{vollmer}. Let
\begin{align*}
p(a)&:=P(X_i=a),\quad q=1-\min_ap(a),\quad p_o=P(X_i= Y_i)=\sum_a
p(a)^2,\\
\p&:=\max_{a\in {\cal A}}p(a),\quad
\q:=1-\min_{a,b}P(X_i=a|Y_i=b),\quad \rho:={p_0\q \over \p
q}.\end{align*} When $X$ and $Y$ are independent, then $\q=q$. The
following theorem is the main result of \cite{vollmer}. Below, it is
stated for vertical sequence, but it also holds for horizontal
distance.  The condition (\ref{maincondition1}) postulates the
relatedness. It can be shown that for independent or very little
related sequences (\ref{maincondition1}) is not fulfilled. It does
not mean that for independent sequences the inequality (\ref{ineq})
fails, but the simulations in \cite{vollmer} show that this is the
case. In the following theorem, $h$ stands for the binary entropy
function, i.e, $h(p)=-p\log_2p-(1-p)\log_2(1-p)$.
%---------
\begin{theorem}\label{thm}
Let $X$ and $Y$ be related. Assume
\begin{equation}\label{maincondition1}
\gamma_{\tt R}\log_2 \p+(1-\gamma_{\tt R})\log_2 (q \q)+
\big((1-\gamma_{\tt R})\wedge \gamma_{\tt R})\log_2\big(\rho \vee
1\big)+ 2h(\gamma_{\tt R })<0.
\end{equation}
Then there exist constants  $C<\infty$ and $D<\infty$ such that for
$n$ big enough,
\begin{equation}\label{ineq}
P\bigl(V_n>C\ln n\bigr)\leq Dn^{-2},
\end{equation}
where $V_n$ is the maximal vertical distance between extremal
alignments. \end{theorem}
%---------------
\section{The case study}
Based on Theorem \ref{thm} as well as the simulation study, we
conjecture that properties of the extremal alignments could be used
as a measure of relatedness. In particular, the maximal horizontal
and vertical distance between extremal alignments might me a good
measure. Also, as one can see from the Figures 1 and 2 (and from
other similar simulations), for independent sequences, there are
relatively long intervals where the  extremal alignments do not
coincide. In Figure 1, the biggest such interval has length 458 (the
end of that interval is marked with *). This interval is called the
{\it maximal non-uniqueness stretch}, and we conjecture that this
can be a good measure of homology as well. A related criterion is
the number of points  where the extremal alignments coincide. We
call it the {\it number of uniqueness points}. Clearly in Figure 1
the number of uniqueness points is relatively small in comparison
with Figure 2.\\\\
%---------------------
We studied four bacterial genes with comparable length (about 1500
letters). The gene is {\it dnaA} and they were taken from bacteria
{\it Pseudomonas putida F1} (Gene nr.1), {\it Pseudomonas syringae
pv. syringae B728a} (Gene nr. 2), {\it Escherichia coli E24377A}
(Gene nr. 3) and {\it Erwinia carotovora subsp. atroseptica
SCRI1043} (Gene nr 4). The corresponding DNA-sequences can be found
in Appendix. All the genes have the same function, therefore they
are presupposed similar. The results of the case study are in the
following table.\\\\
\begin{center}
\begin{tabular}{|c||c||c||c||c|}
  \hline
  Genes & 1 & 2 & 3 & 4 \\\hline\hline
  Max; total  & 2738;  2738 & 1521;  1521 & 625; 671 & 529; 529\\
  Query & 100 & 100 & 71 & 61 \\
  E-value & 0 & 0 & 2e-175 & 1e-146 \\
  MaxIdent & 100 & 82 & 75 & 72 \\\hline
  LCS & 1518 & 1298 & 1081 & 1055 \\
  Vert+Hor=Sum & 0+0=0 & 12+11=23 & 17+18=35 & 20+24=44\\
  non-uniq st. & 0 & 26 & 79 & 111 \\
  uniq points & 1518 & 1003 & 604 & 520 \\
  \hline\hline
  Max; total  & 1521;  1521 & 2771; 2771 & 668; 722 & 538; 592\\
  Query & 100 & 100 & 70 & 69 \\
  E-value & 0 & 0 & 0 & 3e-149 \\
  MaxIdent & 82 & 100 & 76 & 73 \\\hline
  LCS & 1298 & 1536 & 1097 & 1071 \\
  Vert+Hor=Sum & 12+11=23 & 0+0=0 & 15+13=28 & 14+24=38\\
  non-uniq st. & 26 & 0 & 45 & 80 \\
  uniq points & 1003 & 1536 & 633 & 565 \\
  \hline\hline
   Max; total  & 625;  671 & 668; 722 & 2533; 2533 & 1323; 1323\\
  Query & 76 & 76 & 100 & 100 \\
  E-value & 2e-175 & 0 & 0 & 0 \\
  MaxIdent & 75 & 76 & 100 & 81 \\\hline
  LCS & 1081  & 1097 & 1404& 1196 \\
  Vert+Hor=Sum & 17+18=35  & 15+13=28  & 0+0=0 & 6+6=12\\
  non-uniq st. & 79 & 45 & 0 & 21 \\
  uniq points & 604 & 633 & 1404 & 868 \\
  \hline\hline
   Max; total  & 529;  529 & 538; 592 & 1323; 1323 & 2522; 2522\\
  Query & 67 & 76 & 100 & 100 \\
  E-value & 1e-146 & 2e-149 & 0 & 0 \\
  MaxIdent & 72 & 73 & 81 & 100 \\\hline
  LCS & 1055  & 1071 & 1169& 1398 \\
  Vert+Hor=Sum & 20+24=44  & 14+24=38  & 6+6=12 & 0+0=0\\
  non-uniq st. & 111 & 80 & 21 & 0 \\
  uniq points & 520 & 565 & 868 & 1398 \\
  \hline
\end{tabular}
\end{center}
In the table, every (double) cell represents several similarity
criterion between two genes. In the upper part of the cell, the
standard outputs of BLAST-program is represented. The entries "Max"
and "Total" are the maximum and total scores, respectively; "Query"
is the Query-coverage, "E-value" and "MaxIdent" are the e-value and
max-ident, respectively. All parameters of BLAST were deliberately
chosen default. The second half of the cell corresponds to the
extremal alignments-based criterions. "LCS" stands for the length of
the LCS, "Vert+Hor=Sum" is the sum of maximal vertical and
horizontal distance between the extremal alignment, "non-uniq st."
is the length of the longest non-uniqueness stretch and "uniq
points" is the number of uniqueness points of the extremal
alignments.\\\\
From the table, it is evident that Genes 1 and 2  and 3 and 4 are
closely related: the maximum and total scores of BLAST between pairs
(Gene 1, Gene 2) and ( Gene 3, Gene 4) are remarkably higher than
the ones of any other pair of different genes. Note that this
difference  is also well represented by the number of uniqueness
points and, remarkably well by the length of the longest
non-uniqueness stretch. Also, the sums of maximum horizontal and
vertical distances are in full correspondence with  other criterions
measuring well the degree of relatedness. Finally and, perhaps, most
importantly note that all extremal alignments based criterions seem
to be more sensible to  the relatedness, although also  the length
of LCS shows the similarities rather well.

\section{Appendix}
\underline{Gene1: Pseudomonas putida F1}\\\\
{\footnotesize
GTGTCAGTGGAACTTTGGCAGCAGTGCGTGGAGCTTCTGCGCGATGAACTGCCTGCCCAGCAATTCAACA
CCTGGATCCGTCCGCTACAGGTCGAAGCCGAAGGCGACGAGTTGCGCGTCTATGCGCCTAACCGTTTCGT
TCTCGATTGGGTCAATGAAAAGTACCTGGGTCGTTTGCTCGAGCTGTTGGGTGAGAACGGTAGCGGCATT
GCACCAGCCCTTTCCTTATTAATAGGTAGCCGCCGCAGCTCGGCCCCAAGGGCTGCACCCAACGCGCCGG
TCAGCGCTGCCGTTGCGGCTTCGCTGGCGCAGACTCAGGCGCACAAGACGGCCCCGGCAGCAGCGGTTGA
ACCCGTTGCCGTGGCCGCGGCCGAGCCTGTATTGGTCGAGACGTCTTCGCGTGACAGCTTTGATGCCATG
GCCGAGCCTGCTGCTGCGCCGCCCAGTGGTGGCCGGGCTGAACAGCGCACCGTGCAGGTTGAAGGTGCGC
TCAAGCACACCAGTTACCTGAACCGGACCTTTACCTTTGACACCTTCGTCGAAGGTAAGTCGAACCAGCT
CGCCCGCGCGGCTGCCTGGCAGGTTGCGGACAACCCTAAGCATGGCTACAACCCACTGTTCCTTTATGGC
GGTGTGGGTTTGGGTAAAACCCACCTTATGCATGCTGTGGGTAACCATCTGCTGAAGAAGAATCCGAACG
CCAAGGTGGTGTACCTGCATTCGGAGCGCTTCGTCGCGGACATGGTCAAAGCGTTGCAACTCAACGCCAT
CAACGAATTCAAGCGCTTCTACCGCTCGGTGGACGCGTTGCTGATCGACGATATCCAGTTCTTCGCTCGC
AAAGAGCGCTCGCAAGAAGAGTTTTTCCACACCTTCAACGCCTTGCTTGAGGGTGGCCAGCAGGTAATCC
TTACCTCTGACCGCTATCCCAAGGAAATCGAAGGCCTGGAAGAGCGTCTGAAGTCGCGCTTTGGTTGGGG
CCTGACGGTGGCTGTCGAGCCGCCAGAGCTGGAGACCCGCGTAGCGATCCTGATGAAGAAGGCCGACCAG
GCCAAAGTCGAGCTCCCGCATGACGCAGCCTTTTTCATCGCTCAGCGCATCCGGTCCAACGTCCGTGAGC
TGGAAGGTGCACTGAAGCGAGTTATTGCTCACTCGCACTTCATGGGGCGTGACATCACCATCGAGCTGAT
TCGTGAATCGCTCAAGGATCTGTTGGCGCTGCAAGACAAACTGGTCAGTGTGGATAACATTCAGCGTACC
GTCGCTGAGTACTACAAGATCAAGATCTCCGATCTGTTGTCCAAGCGTCGTTCGCGTTCTGTCGCGCGCC
CGCGTCAGGTAGCCATGGCCCTGTCCAAGGAGTTGACCAACCACAGTCTGCCGGAAATCGGCGACATGTT
CGGTGGTCGCGACCATACGACCGTGCTGCACGCCTGCCGCAAAATCAATGAACTGAAGGAATCCGACGCG
GACATCCGCGAGGACTACAAGAACCTGCTGCGGACGCTGACGACCTGA }
\\\\
\underline{Gene2: Pseudomonas syringae pv. syringae B728a}\\\\
{\footnotesize
GTGTCAGTGGAACTTTGGCAGCAGTGCGTGGAGCTTTTGCGCGATGAGCTGCCTGCCCAGCAATTCAACA
CTTGGATCCGTCCGCTACAGGTCGAAGCCGAAGGCGACGAGTTGCGTGTGTACGCACCCAATCGTTTTGT
TCTCGACTGGGTCAACGAAAAGTACCTTGGTCGTCTGCTCGAGCTTCTCGGCGAACACGGTCAAGGCATG
GCCCCTGCTCTTTCCTTATTAATAGGAAGCAAGCGCAGCTCAGCACCGCGTGCTGCCCCGAATGCACCCT
TGGCCGCTGCAGCCTCACAGGCGCTGTCTGCCAATTCGGTCAGCAGCGTCTCGGCCCCGGCTCCTGCCAC
GGCTGCTCCAGCTGCTGCTGTAGCGACGCCTGCACCGGTTCAGAACGTTGCAACACACGACGAACCGTCG
CGTGACAGCTTCGATCCGATGGCCGGAGCCAGCTCGCAACAAGCGCCCGCCCGCGCTGAACAACGTACCG
TCCAGGTAGAAGGTGCGCTCAAGCACACCAGTTACCTGAACCGTACGTTCACGTTCGAAAATTTCGTCGA
GGGTAAGTCCAACCAGCTGGCACGCGCTGCGGCCTGGCAGGTTGCCGACAACCCCAAGCATGGCTACAAC
CCGCTGTTCCTTTATGGCGGCGTGGGTCTTGGTAAAACTCACTTGATGCATGCGGTGGGTAACCACCTGC
TGAAGAAGAACCCGAACGCCAAGGTCGTGTACCTGCATTCGGAGCGCTTCGTTGCAGACATGGTCAAGGC
CTTGCAGCTCAATGCAATCAACGAGTTCAAGCGCTTCTACCGTTCAGTCGATGCGCTGCTGATCGACGAC
ATCCAGTTTTTTGCCCGCAAGGAACGTTCGCAGGAAGAGTTTTTCCACACGTTCAACGCGCTGCTGGAAG
GCGGACAGCAGGTCATTCTGACCAGCGACCGCTATCCCAAGGAAATCGAAGGCCTTGAAGAGCGACTCAA
ATCGCGTTTTGGCTGGGGCCTGACGGTTGCCGTCGAGCCTCCGGAGCTGGAAACCCGCGTGGCGATCCTC
ATGAAAAAAGCAGATCAGGCCAAGGTCGATCTGCCCCATGACGCAGCGTTCTTCATCGCCCAGCGAATTC
GCTCCAACGTCCGTGAGCTGGAAGGTGCGCTCAAGCGCGTCATCGCTCACTCGCACTTCATGGGCCGCGA
CATCACCATCGAGCTGATTCGCGAGTCGCTGAAGGACTTGCTGGCGTTGCAGGACAAGCTGGTCAGTGTG
GATAACATTCAGCGCACTGTCGCCGAGTACTACAAGATCAAGATTTCCGATCTGCTGTCCAAGCGTCGTT
CCCGCTCTGTCGCCCGGCCTCGTCAGGTCGCGATGGCGCTCTCCAAGGAACTCACCAACCACAGTCTTCC
GGAAATCGGTGACGTGTTTGGTGGCCGTGACCACACGACTGTCTTGCACGCATGCCGAAAGATCAACGAG
CTCAAGGAATCCGATGCGGATATCCGCGAGGACTACAAGAACCTGCTGCGCACTCTGACTACGTGA}
%-----------
\\\\
\underline{Gene3: Escherichia coli E24377A}\\\\
{\footnotesize
GTGTCACTTTCGCTTTGGCAGCAGTGTCTTGCCCGATTGCAGGATGAGTTACCAGCCACAGAATTCAGTA
TGTGGATACGCCCATTGCAGGCGGAACTGAGCGATAACACGCTGGCCCTGTACGCGCCAAACCGTTTTGT
CCTCGATTGGGTACGGGACAAGTACCTTAATAATATCAATGGACTGCTAACCAGTTTCTGCGGAGCGGAT
GCCCCACAGCTGCGTTTTGAAGTCGGCACCAAACCGGTGACGCAAACGCCACAAGCGGCAGTGACGAGCA
ACGTCGCGGCCCCTGCACAGGTGGCGCAAACGCAGCCGCAACGTGCTGCGCCTTCTACGCGCTCAGGTTG
GGATAACGTCCCGGCCCCGGCAGAACCGACCTATCGTTCTAACGTAAACGTCAAACACACGTTTGATAAC
TTCGTTGAAGGTAAATCTAACCAACTGGCGCGCGCGGCGGCTCGCCAGGTGGCGGATAACCCTGGCGGTG
CCTATAACCCGTTGTTCCTTTATGGCGGCACGGGTCTGGGTAAAACTCACCTGCTGCATGCGGTGGGTAA
CGGCATTATGGCGCGCAAGCCGAATGCCAAAGTGGTTTATATGCACTCCGAGCGCTTTGTTCAGGACATG
GTTAAAGCCCTGCAAAACAACGCGATCGAAGAGTTTAAACGCTACTACCGTTCCGTAGATGCACTGCTGA
TCGACGATATTCAGTTTTTTGCTAATAAAGAACGATCTCAGGAAGAGTTTTTCCACACCTTCAACGCCCT
GCTGGAAGGTAATCAACAGATCATTCTCACCTCGGATCGCTATCCGAAAGAGATCAACGGCGTTGAGGAT
CGTTTGAAATCCCGCTTCGGTTGGGGACTGACTGTGGCGATCGAACCGCCAGAGCTGGAAACCCGTGTGG
CGATCCTGATGAAAAAGGCCGACGAAAACGACATTCGTTTGCCGGGTGAAGTGGCGTTCTTTATCGCCAA
GCGTCTACGATCTAACGTACGTGAGCTGGAAGGGGCGCTGAACCGCGTCATTGCCAACGCCAACTTTACC
GGAAGGGCGATCACCATCGACTTCGTGCGTGAGGCGCTGCGCGACTTGCTGGCATTGCAGGAAAAACTGG
TCACCATCGACAATATTCAGAAGACGGTGGCGGAGTACTACAAGATCAAAGTTGCGGATCTCCTTTCCAA
GCGTCGATCCCGCTCGGTGGCGCGTCCGCGCCAGATGGCGATGGCGCTGGCGAAAGAGCTGACTAACCAC
AGTCTGCCGGAGATTGGCGATGCGTTTGGTGGTCGTGACCACACGACGGTGCTTCATGCCTGCCGTAAGA
TCGAGCAGTTGCGTGAAGAGAGCCACGATATCAAAGAAGATTTTTCAAATTTAATCAGAACATTGTCATC
GTAA
\\\\
\underline{Gene4: Erwinia carotovora subsp. atroseptica SCRI1043}\\\\
{\footnotesize
GTGTCACTTTCGCTTTGGCAGCAGTGTCTTGCCCGTTTGCAGGATGAGTTACCTGCCACAGAATTCAGTA
TGTGGATACGCCCGTTGCAGGCGGAACTGAGTGATAACACTCTGGCGCTCTACGCCCCCAATCGCTTTGT
GCTGGATTGGGTTCGTGATAAATACTTAAATAATATCAATGTCCTGCTGAATGATTTTTGCGGGATGGAT
GCCCCCTTACTGCGTTTTGAAGTGGGGAGTAAACCGCTGGTTCAAACCATAAGCCAGCCAGCGCAGTCGC
ACCACAACCCTGTCAGCGTTGCACGGCAACAGCCAGTACGCATGGCACCGGTACGCCCAAGCTGGGATAA
CTCGCCTGTACAGGCAGAGCATACCTACCGTTCCAATGTGAACCCGAAACATACGTTTGATAACTTCGTT
GAGGGTAAATCGAACCAGTTAGCACGGGCAGCGGCACGTCAGGTGGCTGACAACCCAGGCGGCGCGTATA
ACCCGCTGTTTCTCTATGGCGGCACTGGCTTGGGTAAAACGCACCTGTTGCATGCAGTGGGGAATGGTAT
TATCGCCCGTAAACCCAACGCGAAGGTGGTCTACATGCACTCCGAGCGTTTCGTGCAGGATATGGTGAAG
GCGTTGCAGAACAATGCGATTGAAGAGTTCAAACGCTACTACCGTTCTGTTGACGCACTGCTGATCGATG
ATATTCAATTCTTCGCTAATAAAGAGCGTTCGCAGGAAGAGTTCTTTCATACCTTTAATGCACTGCTGGA
AGGCAACCAGCAAATCATTCTGACTTCTGACCGCTACCCGAAAGAGATCAATGGTGTGGAAGATCGTCTA
AAATCCCGCTTTGGTTGGGGGTTAACGGTCGCGATTGAACCGCCTGAGCTGGAAACCCGCGTGGCGATTC
TGATGAAAAAGGCAGATGAAAATGACATTCGCTTGCCTGGTGAAGTCGCATTCTTTATTGCTAAACGCCT
GCGTTCTAACGTGCGTGAGTTGGAAGGTGCATTGAACCGCGTTATTGCTAACGCCAATTTTACCGGCCGT
TCGATCACCATTGATTTTGTGCGTGAGGCGCTGCGCGATCTGCTGGCGTTGCAGGAAAAGCTGGTTACTA
TCGACAATATTCAAAAGACCGTGGCGGAATACTATAAAATCAAGATAGCCGACCTGCTGTCTAAACGACG
TTCCCGCTCGGTGGCGCGTCCGCGCCAGATGGCGATGGCGTTGGCGAAAGAACTGACGAATCACAGCCTG
CCGGAAATTGGCGATGCCTTTGGCGGGCGTGATCATACGACGGTGTTGCATGCCTGCCGCAAGATTGAGC
AGTTGCGTGAAGAAAGCCACGACATCAAAGAAGATTTTTCCAATTTAATCAGAACACTATCGTCATAA}


\begin{thebibliography}{999}

\bibitem{vollmer} {Lember, J\"uri, Matzinger, Heinrich and Vollmer, Anna-Lisa
}. {Optimal alignments of longest common subsequence and their path
properties}. (2012), submitted.



\bibitem{Chao} Kun-Mao Chao, Louxin Zhang. {Sequence Comparison: Theory and Methods}.
Springer, (2009).



\bibitem{Durbin}  {Durbin, R. and Eddy, S. and Krogh A. and Mitchison,
G.}. {Biological {S}equence {A}nalysis: {P}robabilistic {M}odels of
{P}roteins and {N}ucleic {A}cids}, {Cambridge University Press},
(1998).


\bibitem{Vingron}  {Waterman, Michael S.\ and Vingron, M.} {Sequence Comparison Significance and {P}oisson Approximation}. {\em Statistical Science},
{\bf 9} (1994), 367--381.

\bibitem{Watermanphase} {Arratia, Richard and Waterman, Michael S.}.
{A phase transition for the score in matching random sequences
              allowing deletions}. {\em Ann.\ Appl.\ Probab.} {\bf
              4} (1994), 200--225.


\bibitem{watermanintrocompbio} Waterman, Michael S. {Introduction to Computational Biology},
{Chapman \& Hall}, (1995).


\bibitem{BLAST} {http://blast.ncbi.nlm.nih.gov/Blast.cgi}

\end{thebibliography}
\end{document}